# Physical-chemical characterization of a PEEK surface irradiated with electron beams

Zh. Alsar[1], B. Rakhadilov[4], N. Almasov[1], N. Serik[1], C. Spitas[1], K. Kostas[1] , *Z. Insepov[1,2,3]
[1] Nazarbayev University, Nur-Sultan, Kazakhstan
[2] National Nuclear Research University (MEPhI), Moscow, Russian Federation
[3] Purdue University, West-Lafayette, IN USA
[4] East Kazakhstan Technical University named after D. Serikbayev, Oskemen, Kazakhstan
* E-mail: zinsepov@purdue.edu

**Keywords:** Electron Beam (EB) Irradiated polymer, polymer degradation, surface change

## Abstract

The changes of surface structure, induced by electron beam irradiation (EBI) on the surface of poly(aryl-ether-ether-ketone) (PEEK), were investigated by using a range of various experimental techniques, such as tribology, Raman, AFM, NMR, FTIR. The topography of the surface layers PEEK irradiated with electron beams was established by X-ray diffraction (XRD). The electron irradiation resulted in degradation of the structure and composition of the irradiated zone was determined and segregation of elements was studied.

## I Introduction

Electron-beam processing of polymers, on the one hand, is one of the promising methods for modifying their surface properties, and, on the other hand, a method for studying the effects of the solar wind particles on polymers used as coating materials for space satellites.

The prospects of surface irradiation methods are due to the possibility of varying the modes of the process, in which the desired change in the surface properties of the material occurs, without changing the bulk properties of the product. Electron beam irradiation of polymer films is mainly used to improve their adhesion to the substrates, as well as to increase mechanical strength, surface hydrophilicity, and elasticity of materials [4, 5, 6, 7].

The use of intense pulsed electron beams makes it possible to change the irradiation parameters: electron energy, beam energy density, pulse duration, influence of the spatial distribution of the released energy and the dynamics of thermal fields in the surface of solids [8, 9]. Under irradiation, the formation of the structure and phase composition of materials is determined by a series of micro- and macro-processes that reflects both, the passage of electrons into a target and energy dissipation.

Low-energy electrons are used to sterilize, cross-link or resurface polymer-based materials. It was found [6, 10] that the treatment of polylactide with electron beams of various energies (tens of keV - units of MeV) leads to a decrease in molecular weight by almost 3 times (for PL after treatment it was $1.2\times10^5$ g/mol), which is associated with the occurrence of *chemical* processes, the rupture of polymer chains in macromolecules, and the rate of dissolution in biological environment after exposure. Changes in these properties can be significant in biomedical applications of materials. It was also shown [11] that electron-beam treatment promotes an increase in density along the depth of the surface layer of a polymer material, associated with cross-linking processes in polymer materials.

Polymeric structures can become a choice in spacecraft production in terms of lightweight and demisability, as far as good thermomechanical properties. In this regards polyether ether ketone (PEEK) is one of a promising materials as a high-performance technopolymer [1]. Poly (aryl-ether-ether-ketone) PEEK is a high-temperature and radiation-resistant thermoplastic material which is used in an increasing number of technical applications [2] , e.g., wire protection in nuclear power stations, production of automotive parts, coating of heat exchangers, medical devices or tubing for purified media [3].

Pulsed linear high-frequency electron accelerators are single-cavity machines operating in the standing half-wave mode. The operating frequencies of the ILU-10 are in the meter radio wave range - 118 MHz. The length of the accelerating gap of the ILU-10 accelerator is 26 cm. The accelerating gap of these machines is shorter than the wavelength in vacuum; therefore, in the process of acceleration, the electrons acquire energy that is practically equal to the maximum voltage on the resonator. The ILU accelerators use a triode electron gun (with a control electrode) located immediately in front of the accelerating gap. The use of a control voltage on an electron gun makes it possible to quickly regulate the beam current and reduce the phase angle of injection, which significantly reduces the spread of the electron energy in the beam [12].

When creating PEEK + graphene composites as coatings for spacecraft, it was interesting to study the effect of radiation on the initial material – PEEK.

## II Experimental part

### 2.1 Electron Beam Irradiation

Electron beam processing (EBP) of polymeric materials was carried out on an industrial pulsed accelerator ILU-10 at the "Park of Nuclear Technologies" JSC (Kurchatov, Kazakhstan).

Table 1. Technical parameters of the industrial pulsed accelerator ILU-10

| Parameters | ILU-10 |
|---|---|
| Energy of electrons, MeV | 2,5–5 |
| Average beam power, kW | 50 |
| Average beam current, mA | 15 |
| Power consumption, kW | 150 |
| Accelerator weight, t | 2,9 |

The modes of electron beam processing of polymeric materials are presented in Table 2. The selected modes of electron beam processing are associated with the design of the accelerator. Taking into account the cumulative effect of EBP on the properties of polymers, the irradiation regimes were also associated with a variation in the irradiation dose, which depends on the number of runs, i.e. from the movement of the irradiated materials relative to the electron beam on a moving table.

Table 2. The modes of electron beam processing of PEEK samples

| Sample | Beam energy, MeV | Beam current, mA | Speed, m/min | Number of runs | Radiation dose for 1 run, kGy | Total radiation dose, kGy |
|---|---|---|---|---|---|---|
| PEEK initial | - | - | - | - | - | - |
| PEEK-1 | 2,7 | 6,84 | 9 | 5 | 10 | 50 |
| PEEK-2 | 2,7 | 6,84 | 9 | 10 | 10 | 100 |
| PEEK-3 | 2,7 | 6,84 | 9 | 20 | 10 | 200 |
| PEEK-4 | 2,7 | 6,84 | 0,8 | 2 | 100 | 200 |
| PEEK-5 | 2,7 | 6,84 | 9 | 30 | 10 | 300 |
| PEEK-6 | 2,7 | 6,84 | 0,8 | 3 | 100 | 300 |
| PEEK-7 | 2,7 | 6,84 | 9 | 40 | 10 | 400 |

### 2.2 Tribological tests

The friction-sliding tribological test was performed on a TRB3 tribometer using the standard ball-and-disk technique. A ball with a diameter of 6.0 mm made of steel - 100Cr6 was used as a counterbody. The tests were carried out at a load of 10 N and a linear velocity of 5 cm / s, a radius of curvature of wear of 2 mm.

The volume of polymer wear after the tribological test was determined using a Profiler 130 (the profile of the wear track was measured).

### 2.3 Microhardness determination

The microhardness was determined using a FISCHER SCOPE HM2000 S measuring system in accordance with the requirements of DIN EN ISO 14577-1. According to the instrument's passport, the hardness measurement range is 0.001–120000 N/mm$^2$, the load setting accuracy is 4 mg, and the displacements are measured with an accuracy of 0.1 nm. The microhardness determination error is 2% of the measured value. The indenter approach speed is 2 μm / s. The range of test loads is 1–2000 mN. The primary processing of the test results was carried out using the software of the WIN-HCU device. A Vickers tetrahedral diamond pyramid with a plane angle of 136 ° was used as an indenter.

### 2.4 Nanoindentation

Nanoindentation was carried out on a NanoScan-4D nanohardness meter in accordance with GOST R 8.748-2011. Young's modulus and hardness were determined by the method of Oliver and Pharr at a load of 10 mN. Nano-identification is a test technique that is the indentation of a rigid tip of a known shape into a material under the action of a gradually increasing load, followed by its removal and registration of the dependence of the tip displacement on the load.

### 2.5 FTIR Spectroscopy

For this purpose, the FTIR Spectrometer Nicolet iS10 (Core Facilities, Nazarbayev University) was used.

### 2.6 Study of surface topography

To assess the effect of ELO on the structural-phase state of PEEK, X-ray phase analysis was carried out on an X'Pert PRO diffractometer (PANalytical, the Netherlands). Shooting modes: diffraction angle from 10 ° to 45 °; scanning step 0.03; exposure time 0.5 s; radiation: Cu Kα; voltage and current: 45 kV and 30 mA. Figure 7 shows diffractograms of PEEK before and after ELO. There are four diffraction peaks with maximum angles of 18.717 °; 20.709 °; 22.677 ° and 28.732 °.

X-ray phase analysis was performed on an X-ray Diffraction (XRD) System - SmartLab (Rigaku).

### 2.7 Raman spectroscopy and AFM

Tests were carried out with help of Raman Spectroscopy & AFM Combined System - LabRAM (Horiba).

## III Results and discussion

### 3.1 Tribological tests

Figure 1 shows graphs of the friction coefficient of PEEK samples before and after ELT.

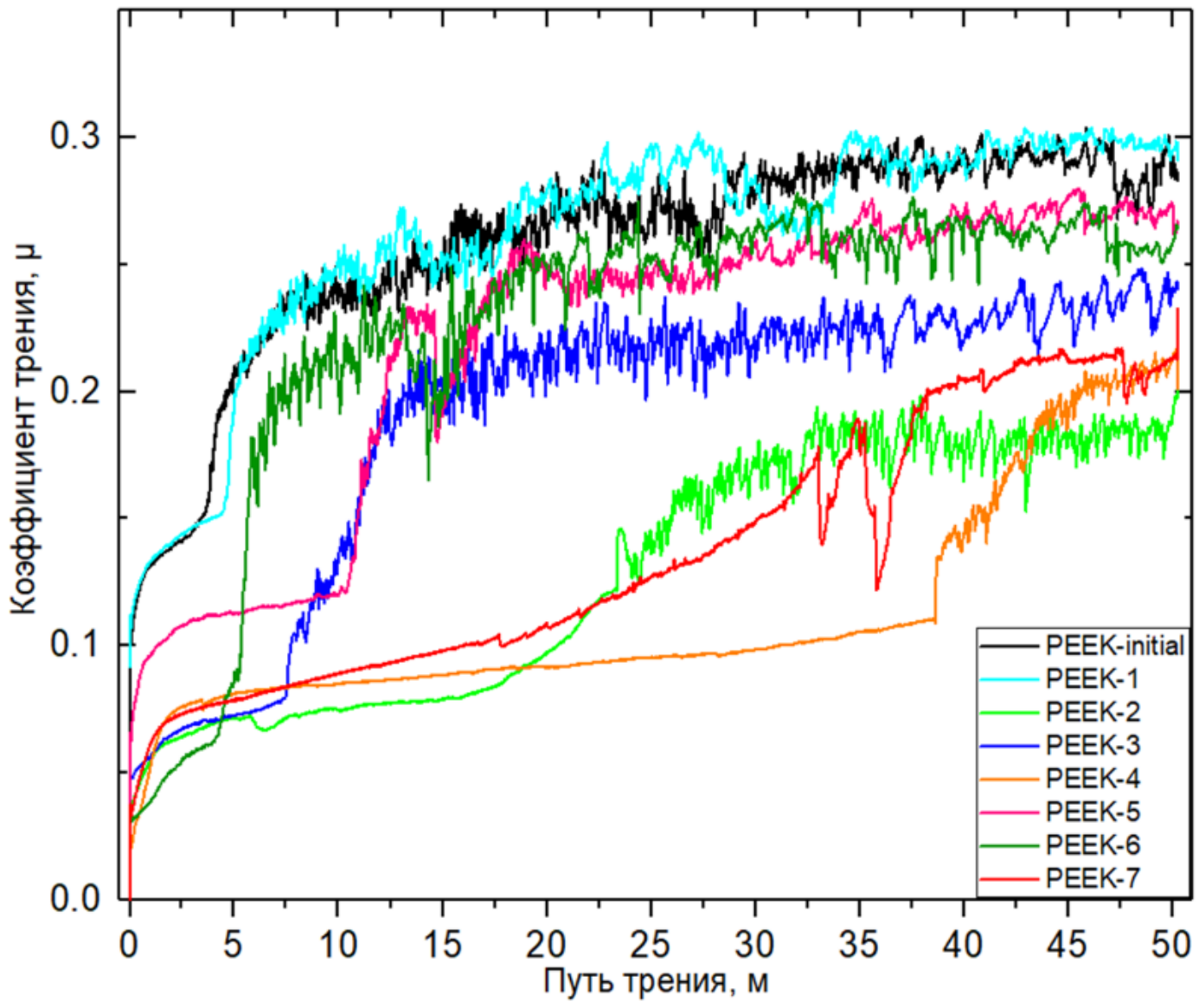

Figure 1 - Graphs of the coefficient of friction of PEEK samples before and after ELO

The tribological characteristics of the samples before and after ELT were characterized by the rate of wear. The wear rate under the action of the tip is calculated based on the volume of material displaced during the test, which was calculated using the following formula (1):

$$I = \frac{V}{F \times l} \quad (1)$$

Where,

I - wear rate, [$mm^3$/N m];

l - Friction path, [m]; F - nominal pressure, [H];

V is the volume of the worn out part, [$mm^3$].

The volume of polymer wear after the tribological test was determined using a Profiler 130 (the profile of the wear track was measured). Figure 2 shows the result of calculating the PEEK wear rate before and after ELO. The results of the study showed a high value of the wear rate of PEEK-1 (radiation dose 50 kGy) in comparison with the original sample. Samples PEEK-4 and PEEK-7 showed high wear resistance.

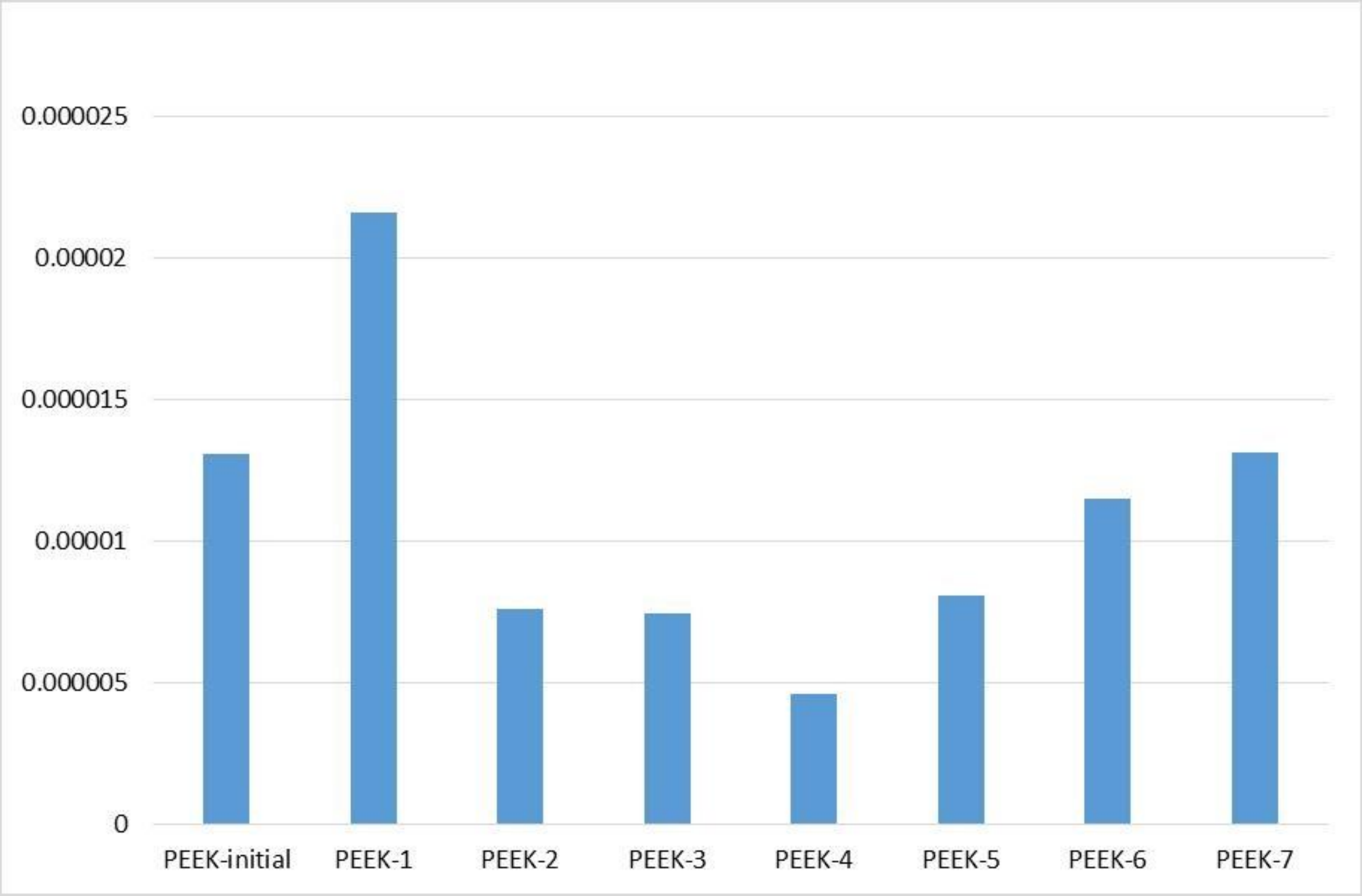

Figure 2 - Intensity of wear of PEEK samples before and after ELO

### 3.2 Microhardness determination

The hardness HM on the Martens scale was chosen as a characterizing parameter. When measuring HM, plastic and elastic deformations are taken into account, so this hardness value can be calculated for a wide range of materials. The Martens hardness is defined as the ratio of the current value of the test load to the cross-sectional area of the indenter at a distance from the apex and is calculated by the formula (2):

$$HM = \frac{F}{A_s(h)} = \frac{F}{26{,}43h^2} \qquad (2)$$

where
HM - Hardness according to Martens, [N / mm$^2$];
F - Test load, [N];
AS - cross-sectional area of the indenter, [mm$^2$];
h - Penetration depth of the indenter, [mm].

Measurements were performed using the traditional load / release method (force increase and decrease). The indenter penetrates into the sample at a maximum force of 500 mN, acting for a specified time (increase in load), which is then removed for a specified time (decrease in force), equal to tn = tp = 20 s.

Figure 3 shows loading diagrams for PEEK specimens before and after ELT. It can be seen that the loading diagrams of the samples coincide without significant displacements (for 5 repeated measurements). This may indicate a uniform effect of ELO on the structure of polymers.

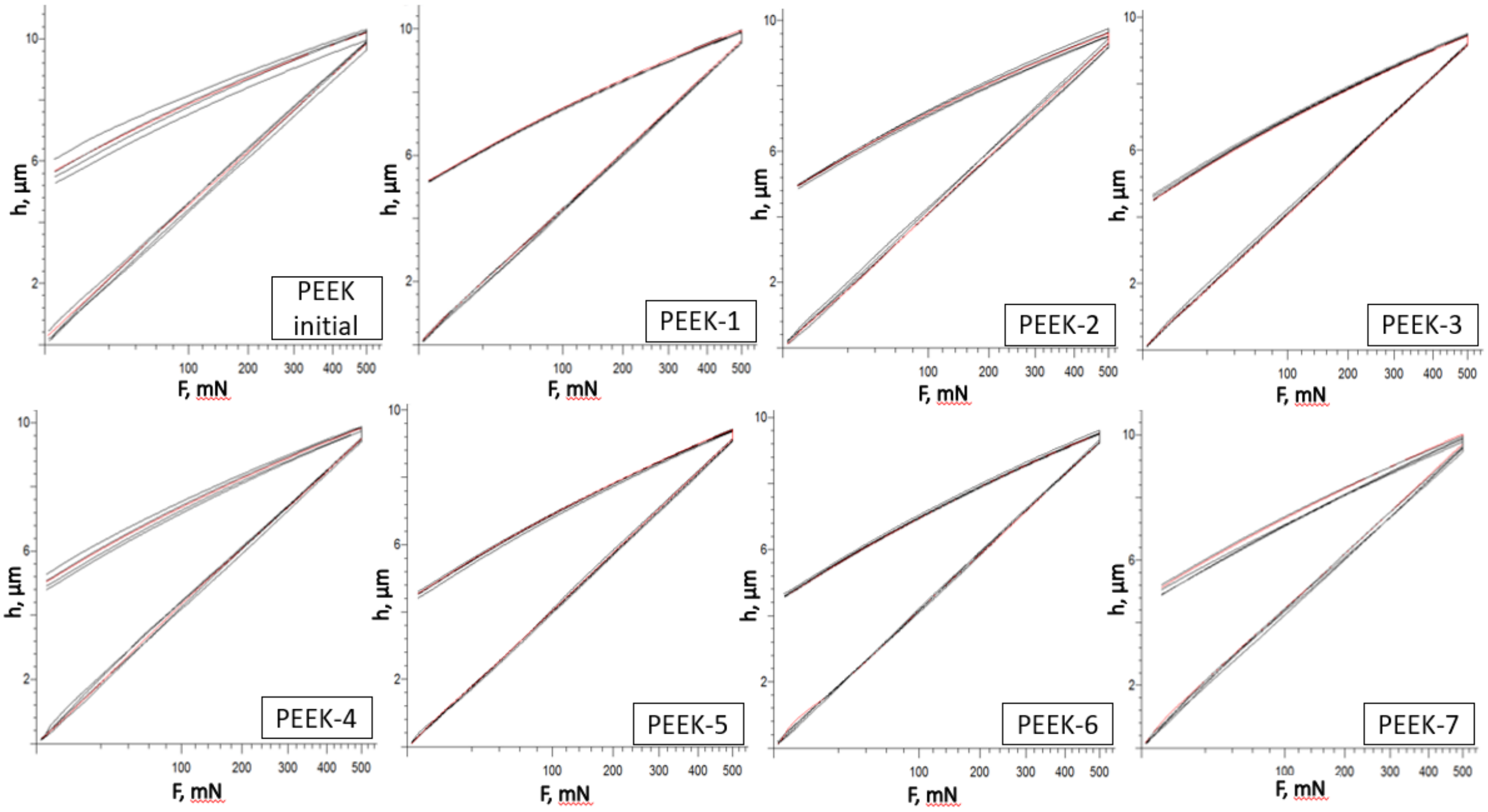

Figure 3 - Graph of the dependence of the indentation depth on the indentation load of PEEK samples, obtained by measuring the hardness according to Martens

Figure 4 shows the results of Martens hardness measurements of PEEK samples before and after ELT. According to the results of the measurements, an increase in the hardness of PEEK after ELO can be observed. The results show the dependence of hardness on the number of runs at the same total irradiation dose (the hardness of PEEK-5 is greater than that of PEEK-6, the hardness of PEEK-3 is greater than that of PEEK-4).

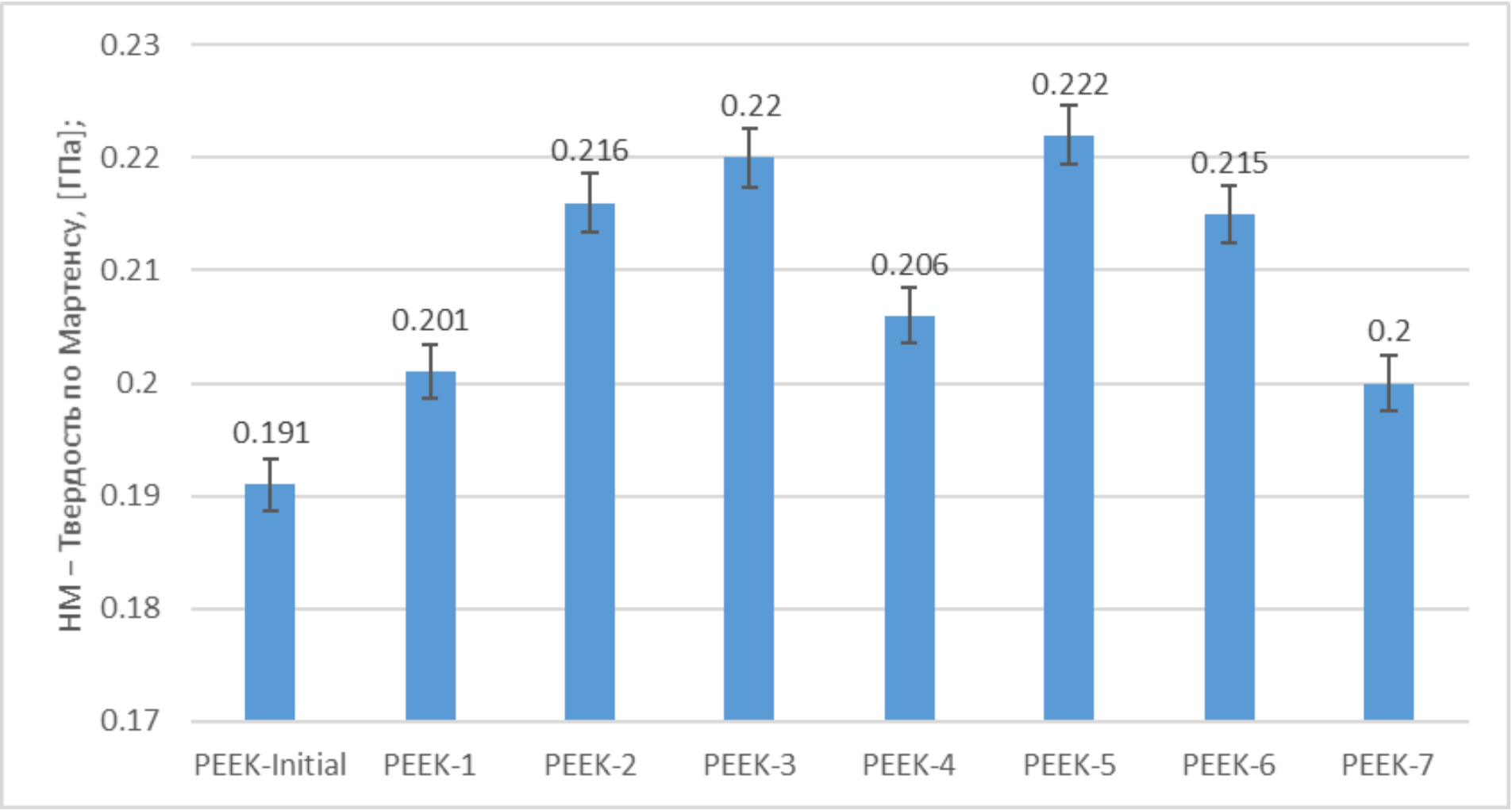

Figure 4 - Results of measuring the hardness of PEEK samples by the Martens method

### 3.3 Nanoindentation

In nanoindentation, the hardness value is calculated by mathematical processing of the diagram (curve) of the dependences of the applied load on the penetration depth of the indenter P (h). The hardness of the sample H is determined from the relation (3):

$$H=\frac{P_{max}}{A_c} \quad (3)$$

where

Pmax - maximum applied load, [mN];

Ac is the area of the contact area, [nm2].

Ac is defined as a function of the contact indentation depth, calculated by the formula (4):

$$h_c = h_{max} - \frac{P_{max}}{s} \qquad (4)$$

where

hc — contact indentation depth, [nm];

hmax - maximum contact indentation depth, [nm];

s is a constant depending on the geometry of the indenter (s = 0.75 for the Berkovich indenter).

The elastic modulus of the test specimen can be obtained from the contact stiffness at the initial moment after removing the load, S = dP / dh, i.e. from the slope of the unloading curve at its beginning. Based on the results of indentation, the value of the contact stiffness can be obtained from the relation (5):

$$S=2\beta \frac{A}{E_r} \qquad (5)$$

where

β is a constant characterizing the shape of the indenter (β = 1.034 for a Berkovich-type indenter);

Er is the reduced modulus of elasticity, which includes the elastic moduli (E, E_i) and Poisson's ratios (v, v_i) of the sample and the indenter, respectively.

Figure 5 shows the results of nanoindentation. The range of indentation depth values is 0.97–1.51 µm for samples after ELT, and for the initial PEEK it is 1.26–1.82 µm.

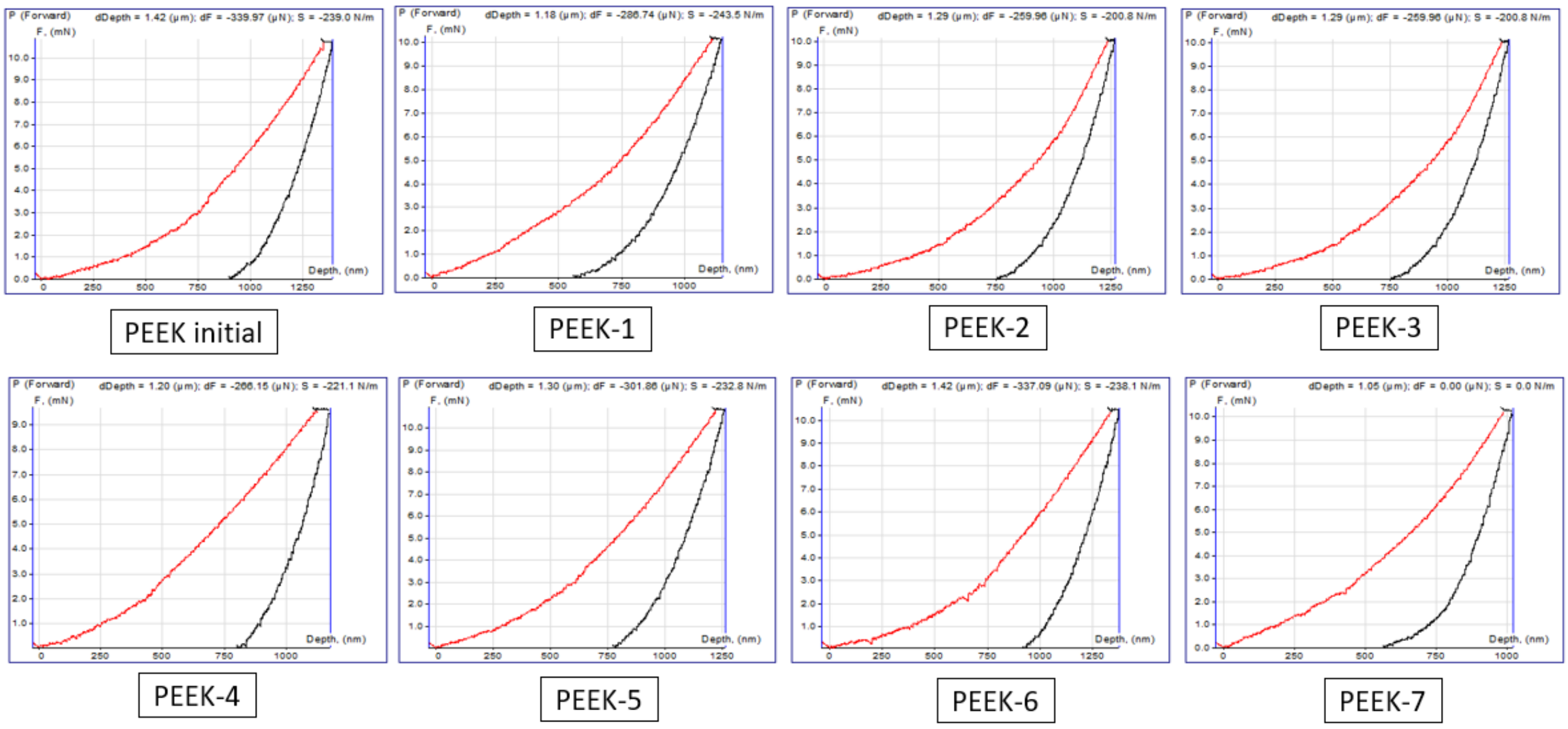

Figure 5 - Graph of the dependence of the indentation depth on the indentation load of PEEK samples obtained by the nanoindentation method

Figure 6 shows the results of measuring the hardness and modulus of elasticity of PEEK samples before and after ELT by the nanoindentation method. According to the results of the measurement, it is also possible to observe an increase in the values of the hardness and elastic modulus of the samples after ELT. Good results were shown by: PEEK-7 (dose of 400 kGy and 40 runs) and PEEK-1 (dose of 50 kGy and 5 runs).

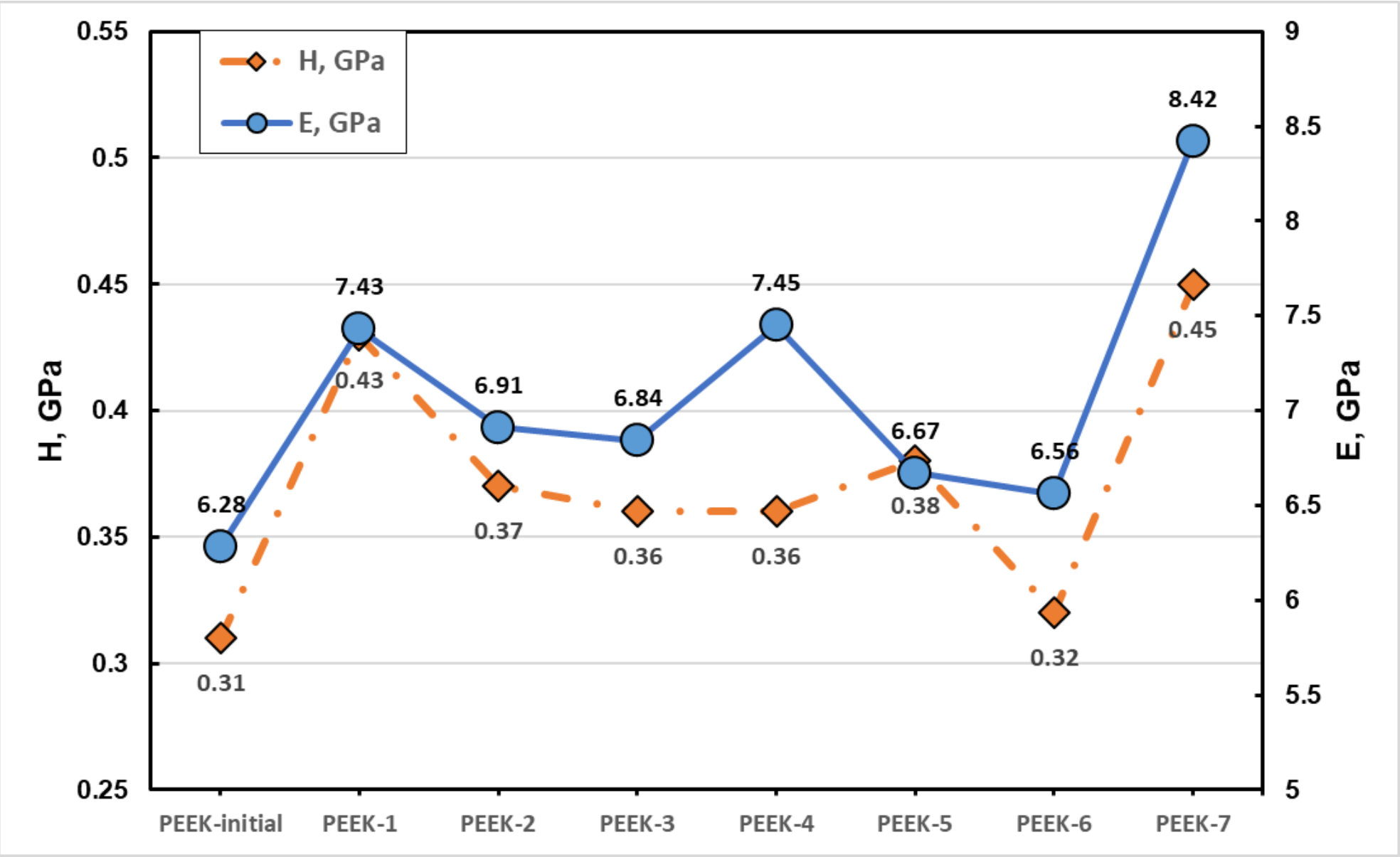

Figure 6 - The results of measuring the hardness and elastic modulus by the nanoindentation method

### 3.4 FTIR spectroscopy

In this section, it was interesting to study the change in the IR spectra of PEEK samples depending on the parameters of electron beam irradiation. The following samples, presented in table 3, were investigated.

Table 3. The modes of electron beam processing of PEEK samples

| Sample | Beam energy, MeV | Beam current, mA | Speed, m/min | Number of runs | Radiation dose for 1 run, kGy | Total radiation dose, kGy |
|---|---|---|---|---|---|---|
| PEEK-0 (initial) | - | - | - | - | - | - |
| PEEK-1 | 2,7 | 6,84 | 9 | 20 | 10 | 200 |
| PEEK-2 | 2,7 | 6,84 | 9 | 10 | 10 | 100 |
| PEEK-3 | 2,7 | 6,84 | 9 | 5 | 10 | 50 |
| PEEK-4 | 2,7 | 6,84 | 0,8 | 2 | 100 | 200 |

Figure 7 shows how irradiation influences the state of chemical bond in PEEK samples in a region of 3200-2700 cm-1.

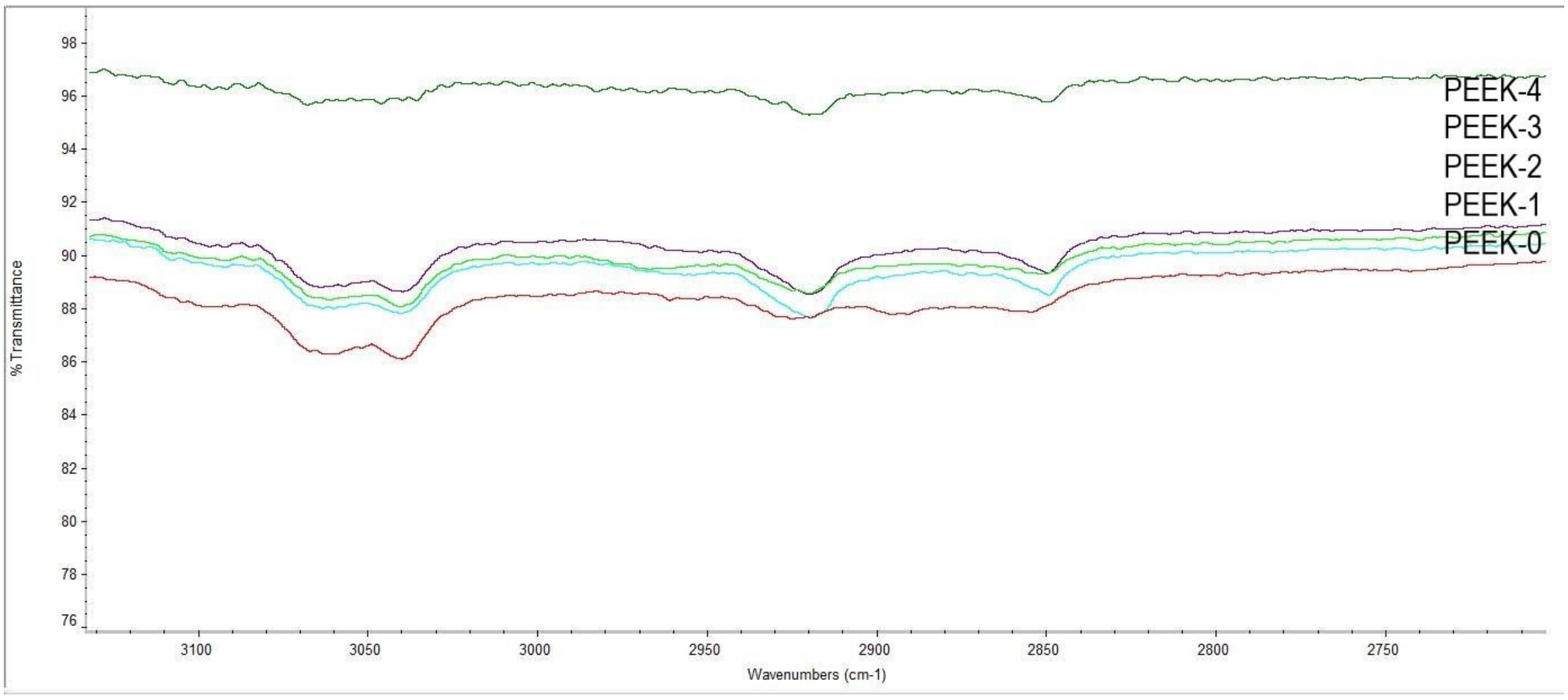

Fig. 7. FTIR spectrum of electron beam irradiated PEEK samples in 3200-2700 cm-1 region

As we can see from the figure 1, the initial PEEK does not show any bond vibrations in 2700-3000 cm-1 region. The range from 2850-3000 $cm^{-1}$ belongs to saturated systems (alkanes, $sp^3$), which are not present in PEEK. However, in irradiated samples we can see some vibrations 2850 cm-1 and 2920 cm-1, which can be an evidence of chemical bonds degradation. The irradiation dose influences the peak intensity. The peaks from 3000-3100 $cm^{-1}$ indicate an unsaturated system (alkenes, $sp^2$, aromatic ring), which are present at the initial PEEK sample and at irradiated samples as well. The intensity of peaks in this region are decreasing in irradiated samples, which also could be an evidence of chemical bonds degradation.

Figure 8 shows how irradiation influences the state of chemical bond in PEEK samples in a region of 1000-600 cm-1.

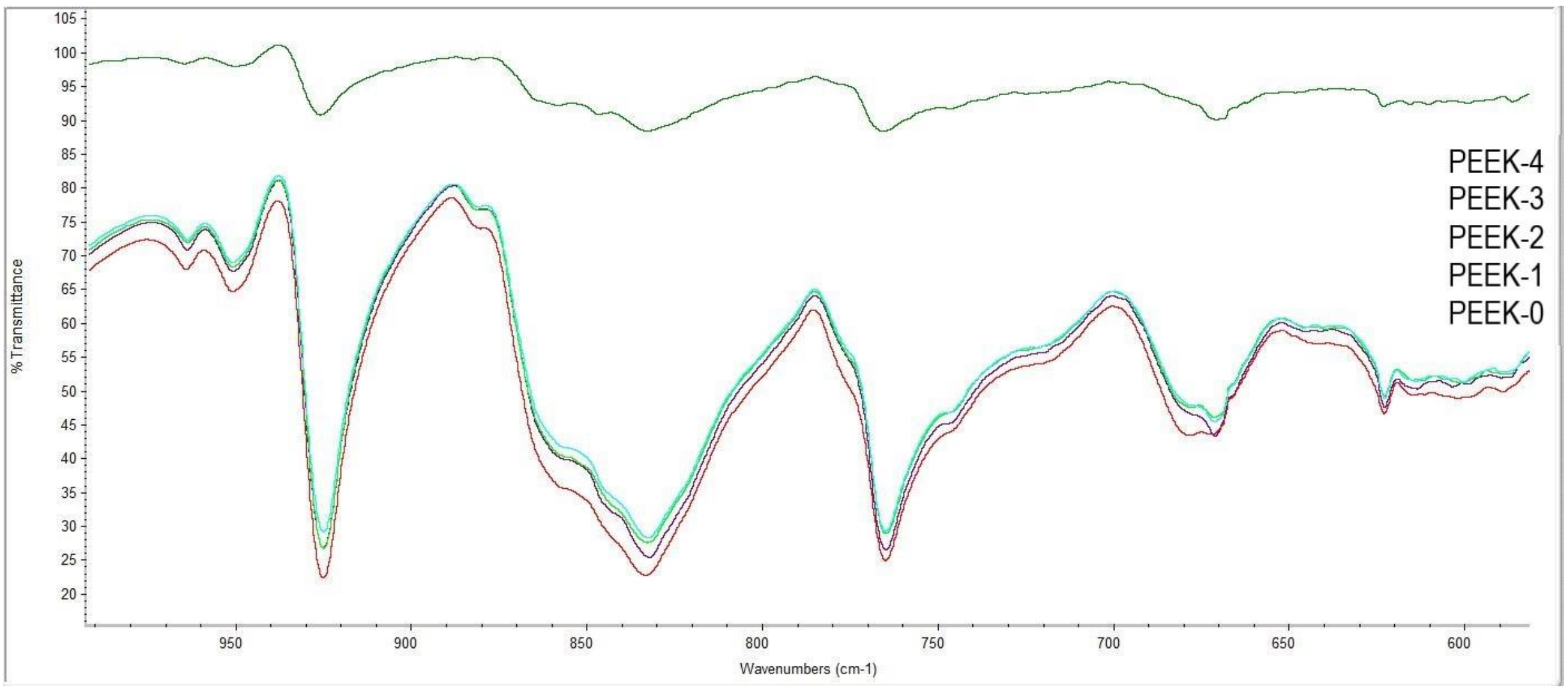

Fig. 8. FTIR spectrum of electron beam irradiated PEEK samples in 1000-600 cm-1 region

On Figure 8, we see the peaks typical for initial PEEK: peaks at 835, 766, 673 cm-1 belong to CH out-plane bending vibration of aromatic ring; the peak at 926 cm-1 belongs to Aryl-(C=0)-Aryl symmetric stretching vibration; the peak at 951 cm-1 refers to aromatic CH out-of-plane for molecules containing at least one complete PEEK repeating unit. In all irradiated samples, we see a decrease in the intensity of the described peaks, which is most noticeable for sample 4 with the highest radiation dose for one run.

We can also see the effect of irradiation on the FTIR spectra of PEEK samples in the region of 1800-1000 cm-1 (Figure 9).

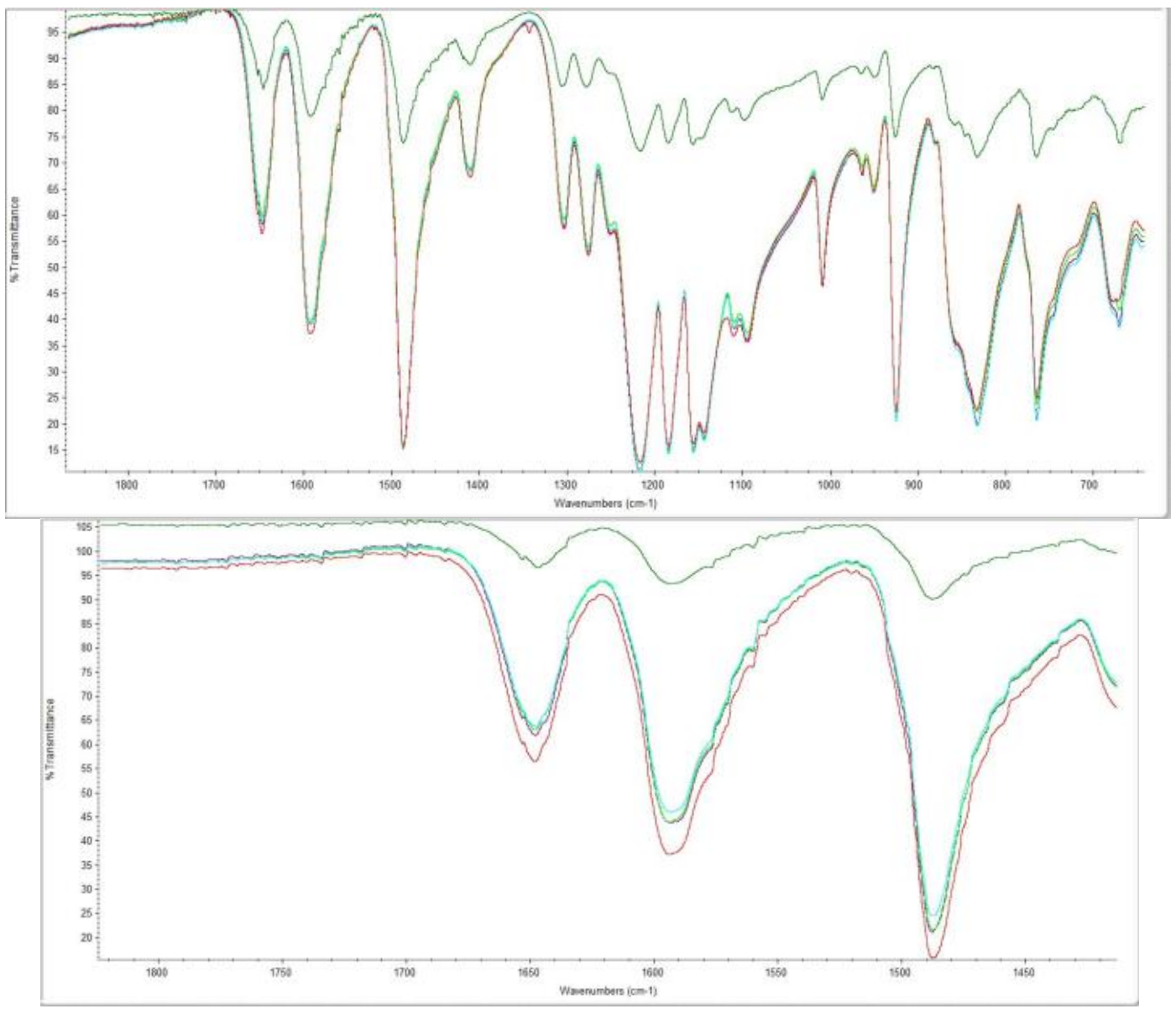

Fig. 9. FTIR spectrum of electron beam irradiated PEEK samples in 1800-1400 cm-1 region

Here peaks 1218, 1185, 1157, 1009 cm-1 indicate C-H in-plane bending vibrations of aromatic ring. Peaks 1593, 1487, 1411 cm-1 refer to benzene skeleton vibrations and 1648 cm-1 - to C=O stretching vibrations. In this area, we again see a decrease in the intensity of the peaks and their actual disappearance with an increase in the radiation dose.

Thus, irradiation of PEEK with an electron beam leads to changes in chemical bonds, to their supposed destruction, as can be seen from the FTIR spectra. The proposed link breaking mechanism is shown on Figure 10.

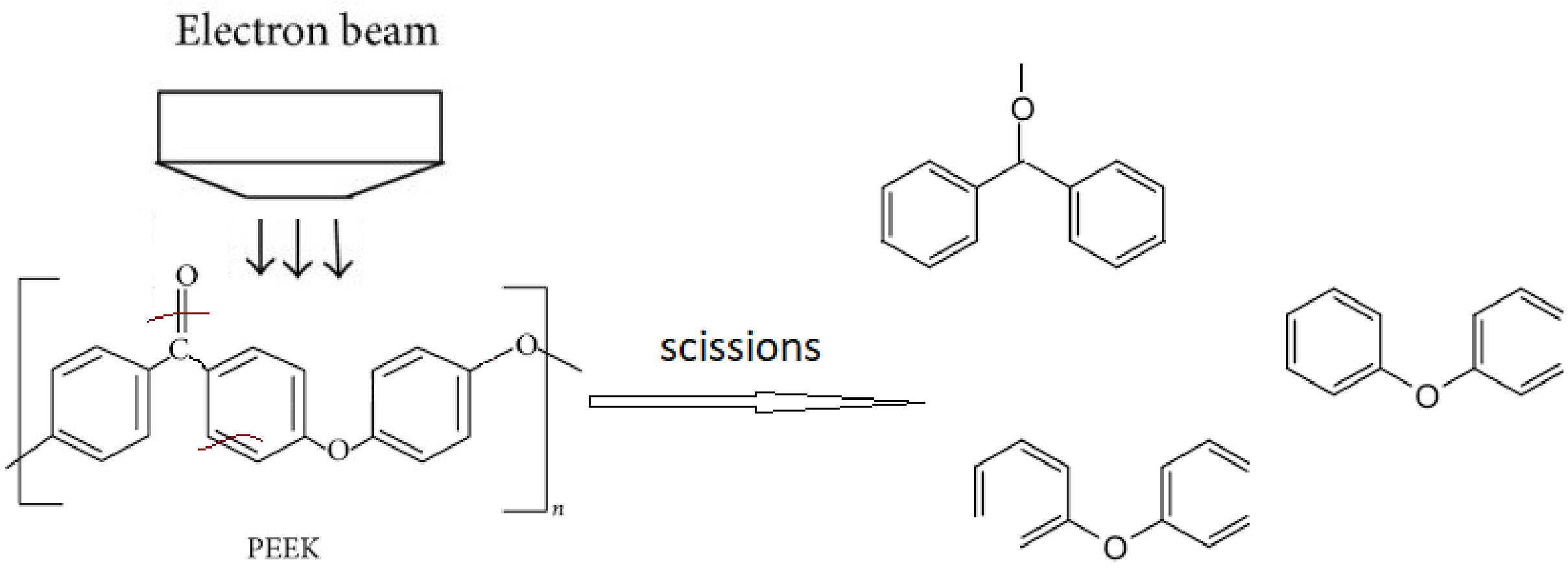

Figure 10. Presumptive mechanism for bonds breaking in PEEK

Currently, there are some published works on the irradiation of PEEK with ion beams and UV radiation, which also leads to the destruction of some bonds, and, mainly, depolymerization of the sample. In our case, when irradiated with an electron beam, electrons presumably act on pi-systems of double bonds and aromatic systems, disrupting their stability, which leads to displacements of the electron density, breaking bonds, and partial depolymerization, which is demonstrated by the FTIR spectra.

### 3.5 Surface topography

To assess the effect of ELO on the structural-phase state of PEEK, X-ray phase analysis was carried out on an X'Pert PRO diffractometer (PANalytical, the Netherlands). Shooting modes: diffraction angle from 10° to 45°; scanning step 0.03; exposure time 0.5 s; radiation: Cu Kα; voltage and current: 45 kV and 30 mA. Figure 11 shows the PEEK diffraction patterns before and after ELO. There are four diffraction peaks with maximum angles of 18.717°; 20.709°; 22.677° and 28.732°. The diffraction patterns at different irradiation doses are practically the same, which indicates the minimal effect of ELO on the crystal structure of PEEK.

There are slight differences in the height and width of the diffraction peaks at different irradiation doses, which suggests a change in the degree of crystallinity (ratio of amorphous to crystalline phases) of PEEK. In work [1] it is reported that crystallinity has a great influence on the tribological and mechanical characteristics of polymers. XRD patterns shown on PEEK are a semi-crystalline thermoplastic phase of a rhombic crystalline form [2]. The vertices around the 2θ diffraction angle of 20° could be indexed as the PEEK planes (110), (113) and (200), respectively. The peak around the 2θ diffraction angle of 29° could be assigned to the PEEK plane (213).

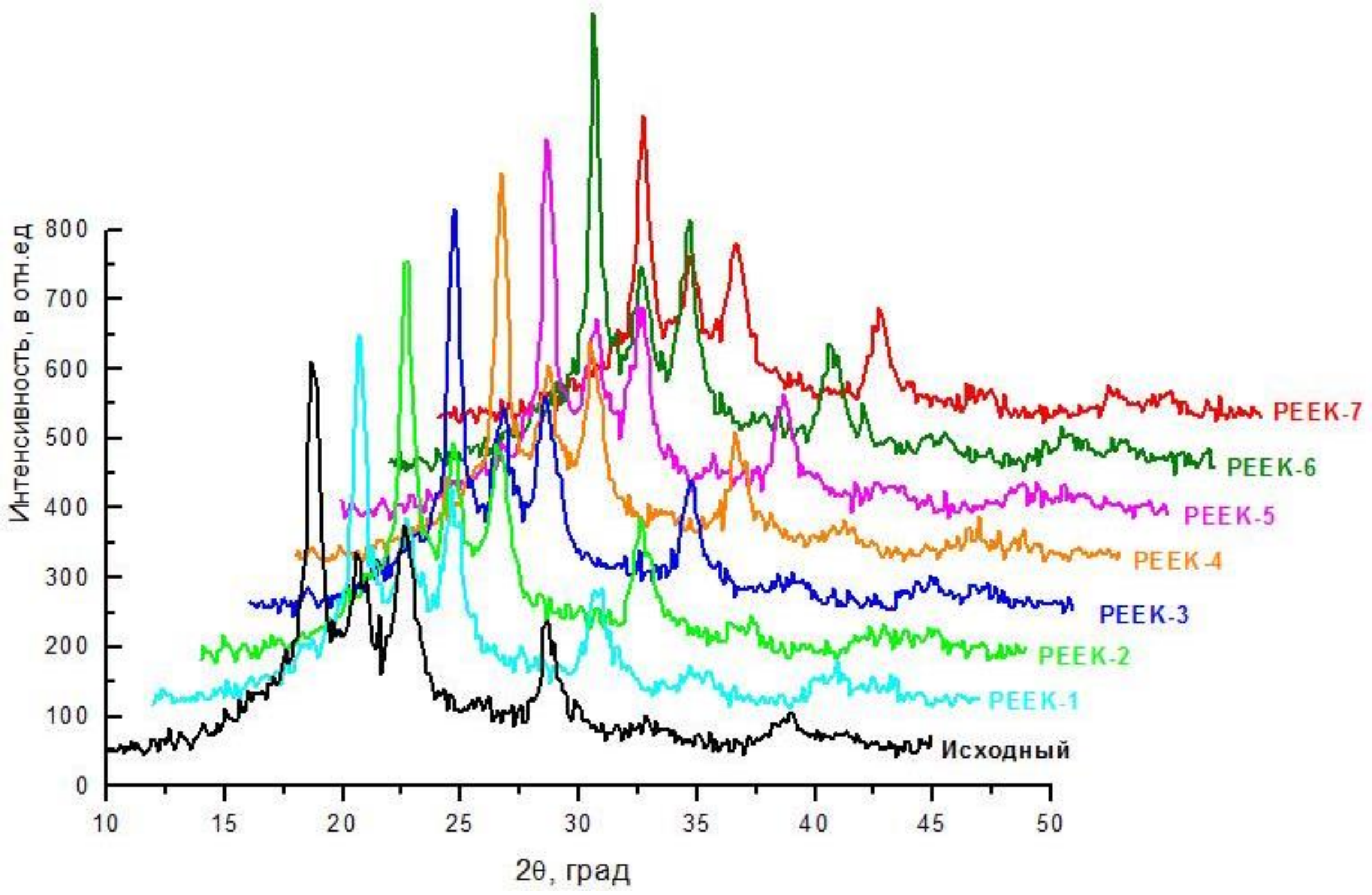

Figure 11 - Diffraction pattern of PEEK polymer before and after ELO

To understand the effect of ELO on the structural-phase state of PEEK, X-ray phase analysis was carried out on an X-ray Diffraction (XRD) System - SmartLab (Rigaku) diffractometer. XRD analysis of a sample of the original PEEK is shown in the figure.

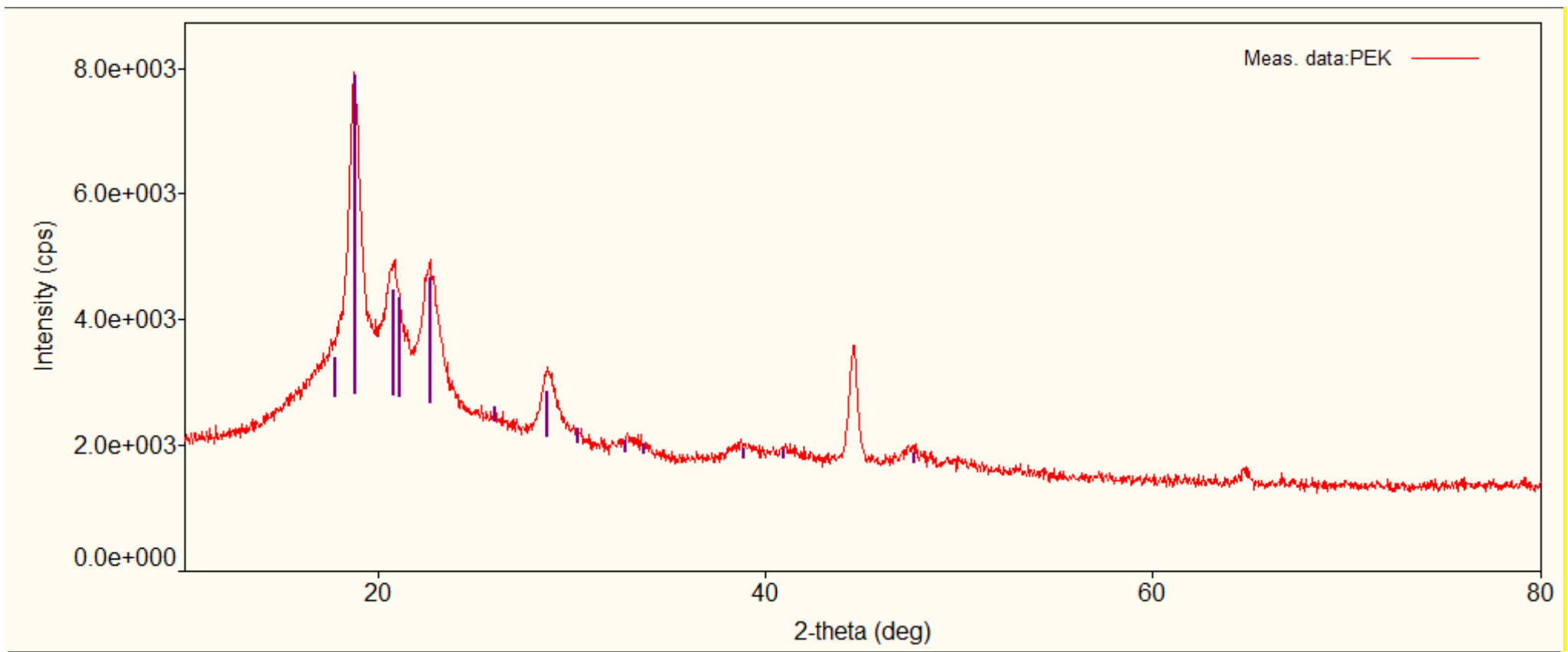

| No. | 2-theta (deg) | Height (cps) | FWHM (deg) | Size (ang.) | Phase name | Chemical formula | Peak shape |
|---|---|---|---|---|---|---|---|
| 1 | 18.768(11) | 2935(66) | 0.683(19) | 123(3) | Unknown | Unknown | Split pseudo-Voigt |
| 2 | 20.85(3) | 1169(41) | 1.30(6) | 65(3) | Unknown | Unknown | Split pseudo-Voigt |
| 3 | 22.63(2) | 1418(46) | 1.04(3) | 81(3) | Unknown | Unknown | Split pseudo-Voigt |
| 4 | 28.77(3) | 672(31) | 0.92(4) | 93(4) | Unknown | Unknown | Split pseudo-Voigt |
| 5 | 33.14(6) | 110(13) | 1.16(15) | 74(10) | Unknown | Unknown | Split pseudo-Voigt |
| 6 | 38.5(2) | 96(12) | 2.4(2) | 37(4) | Unknown | Unknown | Split pseudo-Voigt |
| 7 | 47.48(13) | 133(14) | 1.00(18) | 90(16) | Unknown | Unknown | Split pseudo-Voigt |

Figure 12 - Diffractogramm of PEEK polymer (Initial)

XRD patterns shown on PEEK are a semi-crystalline thermoplastic phase of a rhombic crystalline form [27]. The vertices around the 2θ diffraction angle of 20 ° could be indexed as the PEEK planes (110), (113) and (200), respectively. The peak around the 2θ diffraction angle of 29 ° could be assigned to the (213) PEEK plane.

### 3.6 Raman spectroscopy and AFM

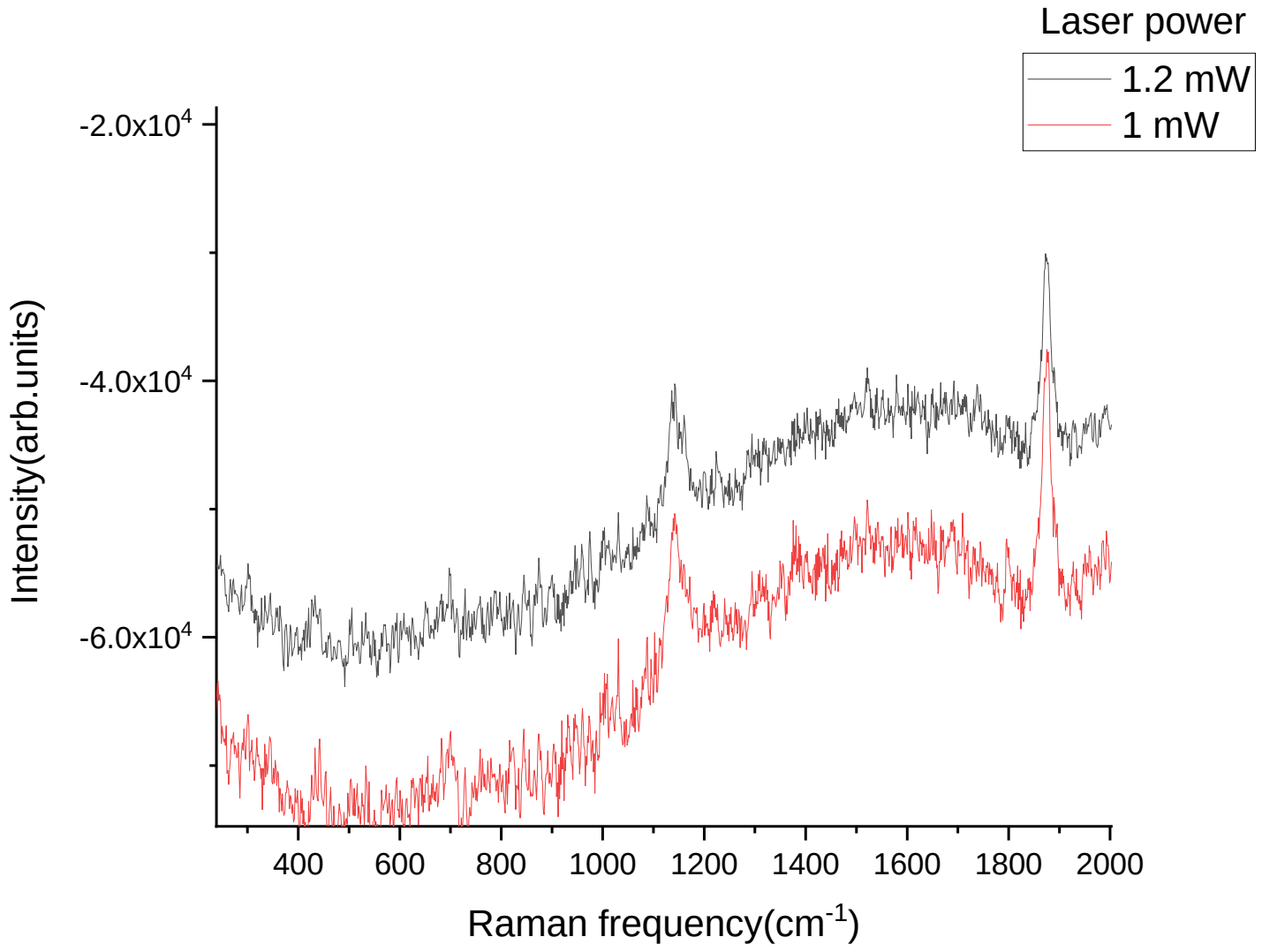

Laser wavelength: 633nm
Grating: 600 gr/mm
Spectral resolution: 5 cm^(-1)
Beam diameter: 2 micron

**Fig. 13 - Raman spectra of PEEK**

The PEEK sample is very luminescent, the luminescence overlaps the Raman signal, where after complex measurement and subtraction processes, very weak Raman peaks are obtained. The intensity values are negative due to the subtraction of the spectra.

AFM images were acquired with an Atomic Force Microscope SmartSPM 1000. An SPM Probe canteliver was used - part

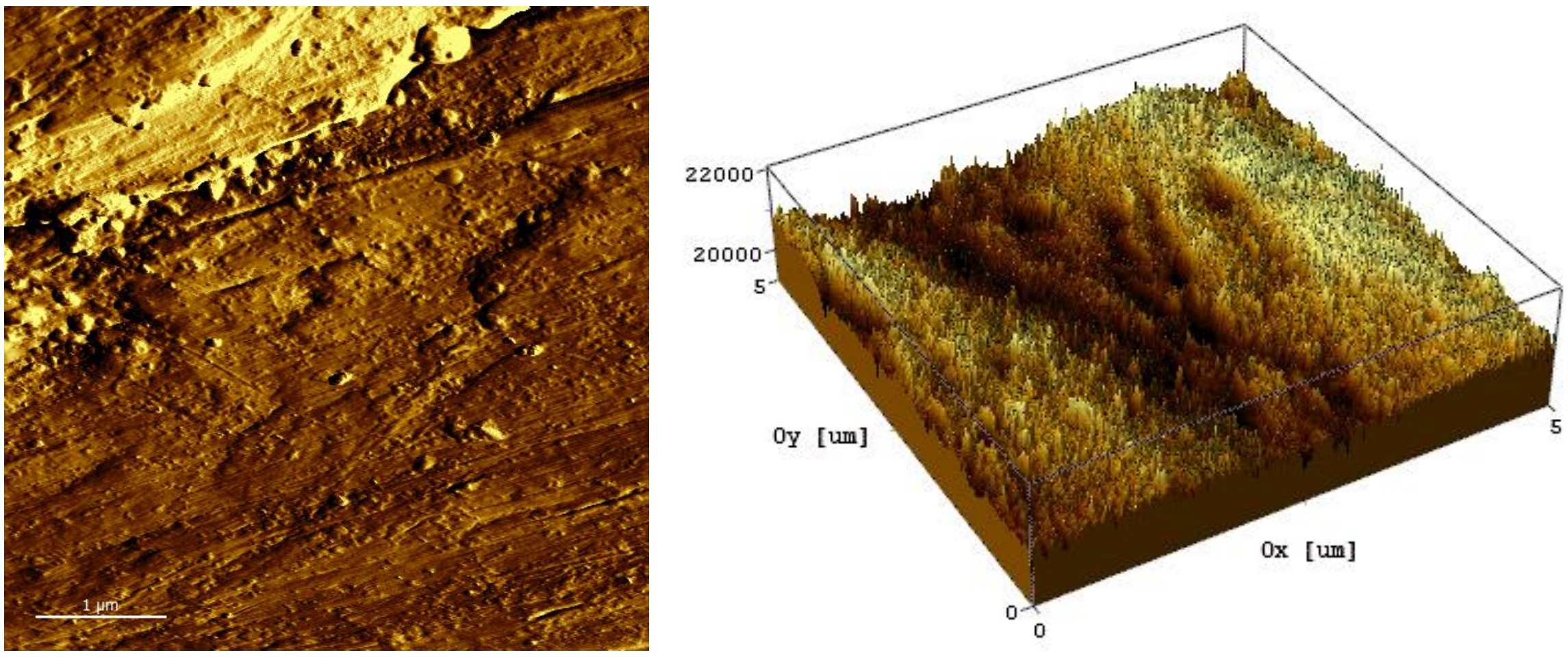

**Fig.14 – AFM image of initial PEEK**

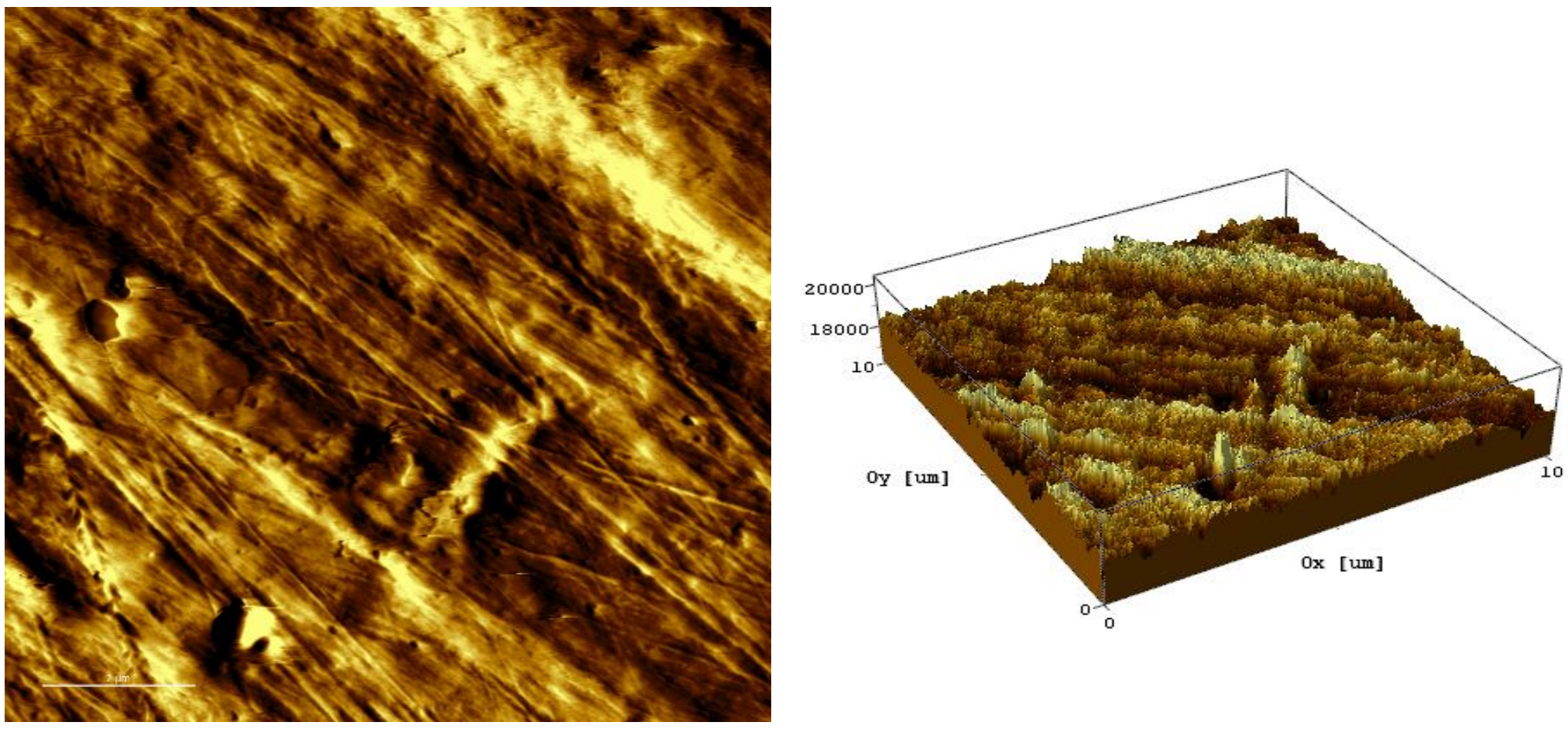

Amount = 1048576
Average value = 5.00099 µm
Maximum = 5.2631 µm
Minimum = 4.78684 µm
Median = 5.01083 µm
Ra = 0.0928808 µm
Rms = 0.106265 µm
Skew = 0.0846833
Kurtosis = -1.11311
Surface area = 102.749 µm²
Projected Area = 101.841 µm²
Inclination θ = 1.79355°
Inclination φ = -147.988°

**Fig.15 – AFM image of PEEK-1**

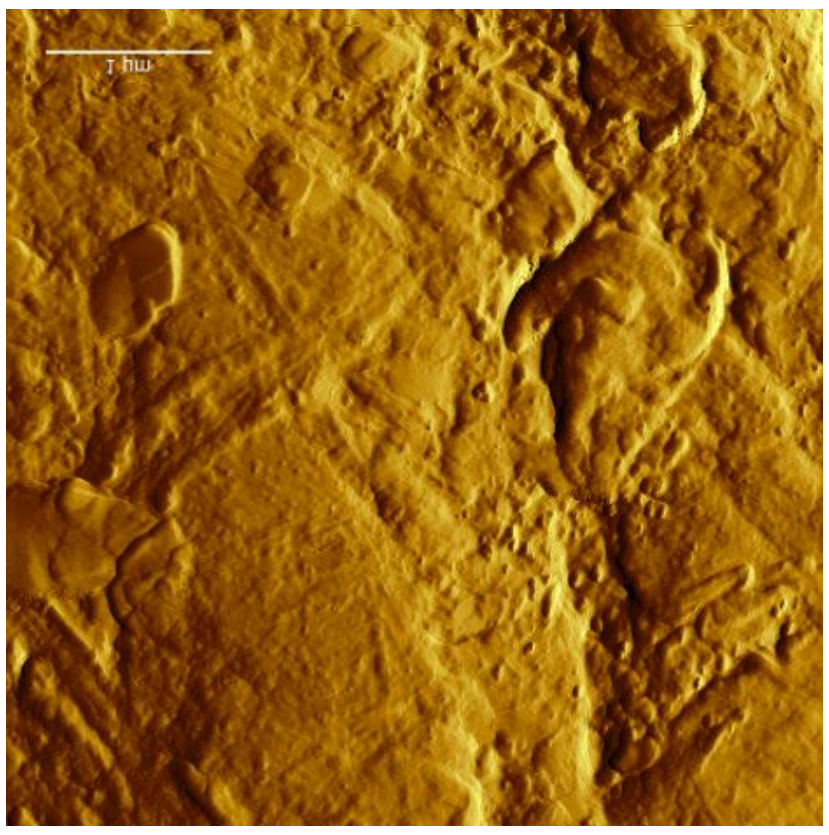

Amount = 262144
Average value = 3.78703 µm
Maximum = 4.01745 µm
Minimum = 3.41955 µm
Median = 3.7996 µm
Ra = 0.0937676 µm
Rms = 0.112255 µm
Skew = -0.0970095
Kurtosis = -0.70157
Surface area = 26.6158 $\mu m^2$
Projected Area = 24.9024 $\mu m^2$
Inclination θ = 2.19115°
Inclination φ = 44.3472°

**Fig.16 – AFM image of PEEK-3**

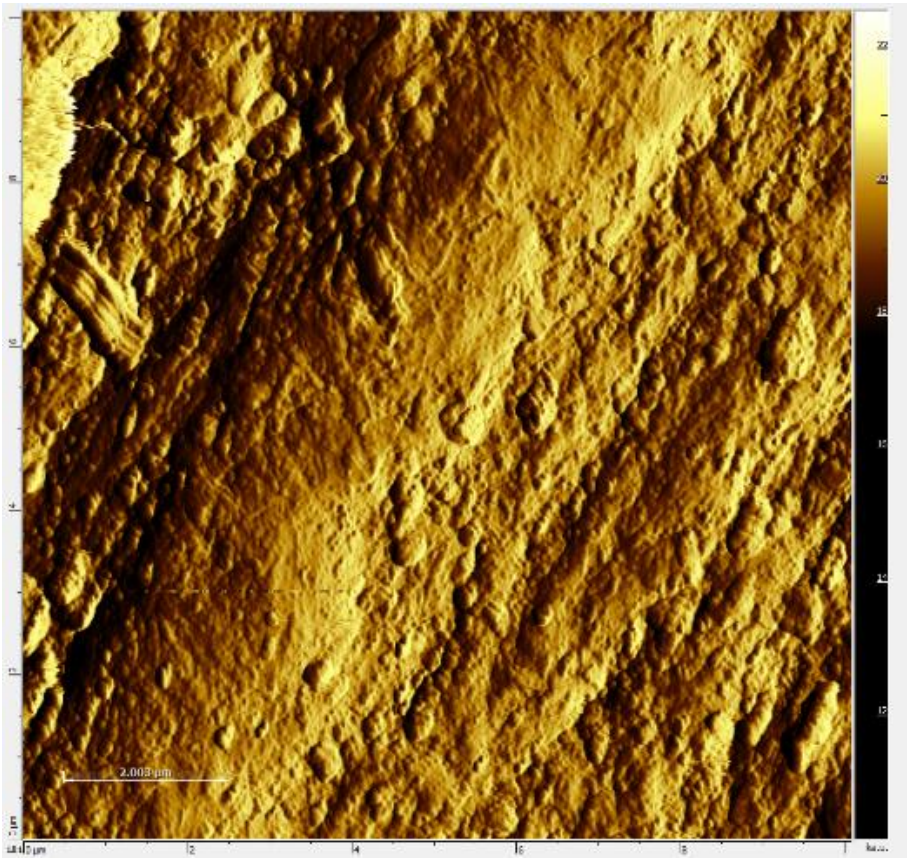

Amount = 262144
Average value = 5.5378 µm
Maximum = 5.97817 µm
Minimum = 4.99064 µm
Median = 5.52489 µm
Ra = 0.156039 µm
Rms = 0.191692 µm
Skew = 0.246249
Kurtosis = -0.702803
Surface area = 109.346 $\mu m^2$
Projected Area = 101.642 $\mu m^2$
Inclination θ = 1.97938°
Inclination φ = -99.6952°

**Fig.17 – AFM image of PEEK-4**

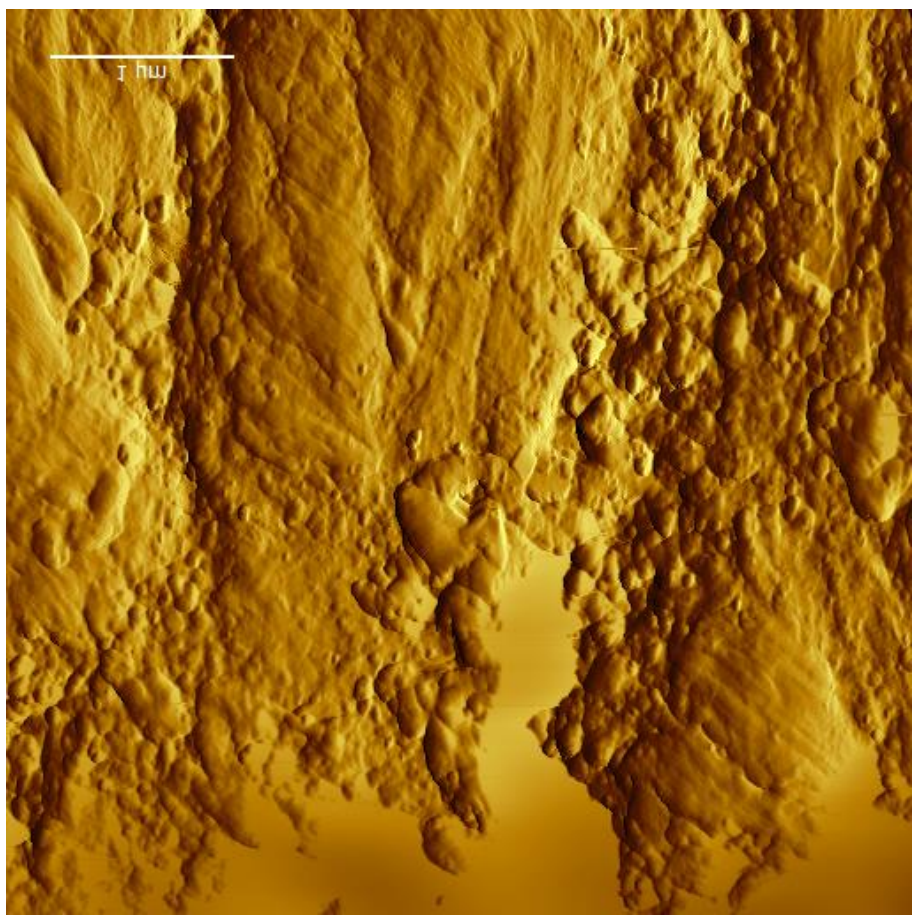

Amount = 262144
Average value = 4.96436 µm
Maximum = 5.14759 µm
Minimum = 4.63726 µm
Median = 4.97949 µm
Ra = 0.0738782 µm
Rms = 0.0928156 µm
Skew = -0.726291
Kurtosis = 0.0360467
Surface area = 26.453 µm²
Projected Area = 24.9024 µm²
Inclination θ = 3.00357°
Inclination φ = -60.9858°

**Fig.18 – AFM image of PEEK-5**

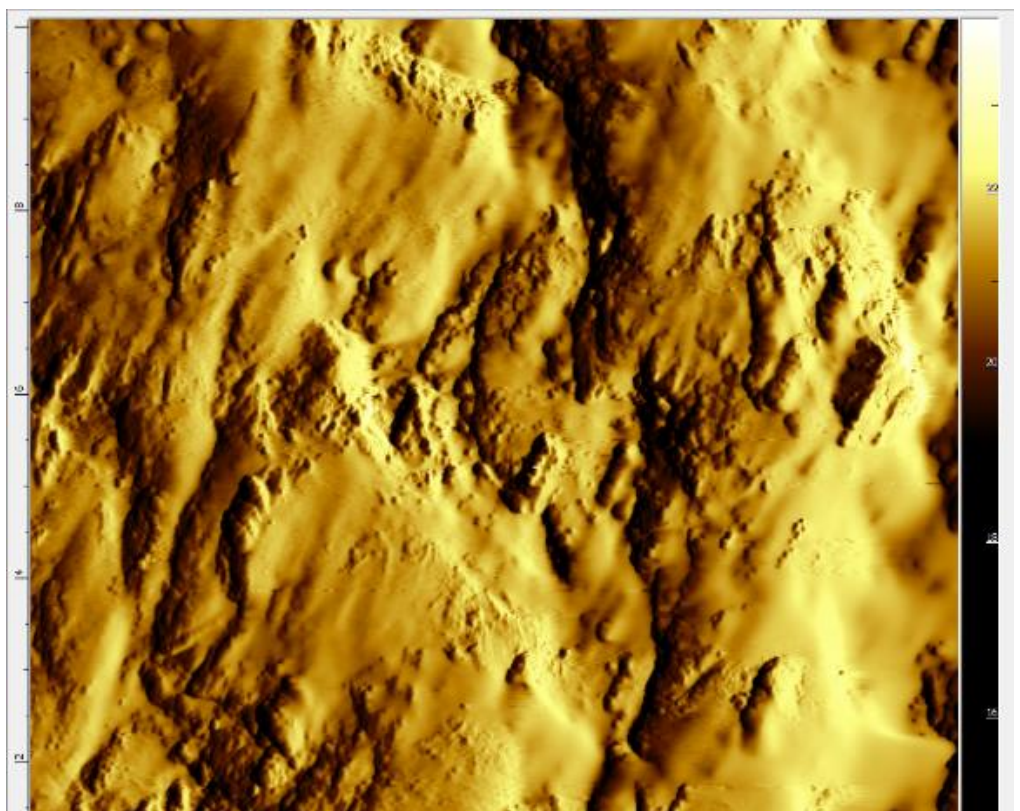

Amount = 262144
Average value = 21.0282 ka.u.
Maximum = 23.986 ka.u.
Minimum = 13.402 ka.u.
Median = 21.195 ka.u.
Ra = 0.665298 ka.u.
Rms = 0.826348 ka.u.
Skew = -0.73603
Kurtosis = 0.234798
Projected Area = 101.642 µm²
Inclination φ = 178.411°

**Fig.19 – AFM image of PEEK-7**

**Acknowledgements:** This research was funded by the Nazarbayev University Collaborative Research Project (CRP): "Development of smart passive-active multiscale composite structure for earth Remote Sensing Satellites (RSS) of ultrahigh resolution (ULTRASAT)", Grant Award Nr. 091019CRP2115.